\documentclass[pra,twocolumn,tightenlines,showpacs]{revtex4}

\usepackage[active]{srcltx} % allows inverse search in Linux
\usepackage{bm,dcolumn,amsmath,graphicx}

\newcommand{\tref}[1]{Table~\ref{#1}}

\begin{document}

\title{Relativistic many-body calculation of energies, lifetimes,  polarizabilities, blackbody radiative shift and hyperfine constants in  Lu$^{2+}$}

\author{U. I. Safronova}
\affiliation {Physics Department, University of Nevada, Reno,
Nevada 89557}
\author{M. S.  Safronova}
\affiliation {Department of Physics and Astronomy, 217 Sharp
Lab, University of Delaware, Newark, Delaware 19716 and
Joint Quantum Institute, NIST and the University of Maryland, College Park, Maryland, USA}
\author{W. R. Johnson}
\affiliation{Department of Physics,  225 Nieuwland Science Hall, University of Notre Dame, Notre Dame, IN 46556}

\date{\today}
\begin{abstract}
Energy levels of 30 low-lying states of Lu$^{2+}$ and allowed electric-dipole matrix elements  between these states
are evaluated  using a relativistic all-order
method in which all single, double and partial triple excitations
of Dirac-Fock wave functions are included to all orders of
perturbation theory. Matrix elements are critically evaluated for their accuracy and recommended values of the matrix elements are given
together with uncertainty estimates. Line strengths, transition rates and lifetimes of the metastable $5d_{3/2}$ and $5d_{5/2}$ states are calculated.  Recommended values are given for static polarizabilities of
the $6s$, $5d$ and $6p$ states and  tensor polarizabilities of the $5d$ and $6p_{3/2}$ states.  Uncertainties of the polarizability values are estimated in all cases.
The blackbody radiation shift of the $6s_{1/2}-5d_{5/2}$  transition frequency of the Lu$^{2+}$ ion is calculated with the aid of the recommended scalar polarizabilities of the $6s_{1/2}$ and $5d_{5/2}$ states.
Finally, $A$ and $B$ hyperfine constants are determined for states of $^{175}$Lu$^{2+}$ with $n \le 9$. This work provides recommended values of transition matrix elements, polarizabilities and hyperfine constants of Lu$^{2+}$, critically evaluated for accuracy, for benchmark tests of high-precision theoretical methodology  and  planning of future experiments.
\end{abstract}
%31.15.A- Ab initio calculations
%31.15.ac High-precision calculations for few-electron (or few-body) atomic systems
%31.15.ag Excitation energies and lifetimes; oscillator strengths
%31.15.aj Relativistic corrections, spin-orbit effects, fine structure; hyperfine structure
%31.15.am Relativistic configuration interaction (CI) and many-body perturbation calculations
%31.15.ap Polarizabilities and other atomic and molecular properties
%31.15.V- Electron correlation calculations for atoms, ions and molecules
%31.15.vj Electron correlation calculations for atoms and ions: excited states
%31.30.J- Relativistic and quantum electrodynamic (QED)
%effects in atoms, molecules, and ions
\maketitle
%\end{document}

\section{Introduction}

The development of high-precision optical atomic clocks is important for many applications \cite{LudBoyYe15},
including precision timekeeping, tracking of deep-space probes, study of many-body quantum systems \cite{ZhaBisBro14},
relativistic geodesy with potential future application to monitor volcanic magma chambers and earthquake prediction \cite{BonBonJet15}, and tests of fundamental physics such as search for variation of fundamental constants \cite{GodNisJon14,HunLipTam14}
tests of Lorentz invariance \cite{WolChaBiz06}, search for topological dark matter \cite{DerPos14}.

While trapped ion clocks achieved ultra-low uncertainty of $3\times10^{-18}$, realised with octupole
transition in Yb$^{+}$ \cite{HunSanLip16}, a significant bottleneck for further improvement of
accuracy arises from  relatively low stability  achievable with a single
ion \cite{LudBoyYe15}. Recently, singly ionized lutetium has been proposed as a
possible candidate to overcome this hurdle via the use of large ion crystals with a special scheme to cancel the effect of micromotion \cite{ArnHajPae15,PaeArnHaj16}. The crucial condition for the implementation of such a scheme is the negative value of the scalar polarizability
difference for the clock transition  \cite{MadDubZho12,DubMadZho13,ArnHajPae15,PaeArnHaj16}.
We find that Lu$^{2+}$ is another system which satisfies such a condition with low-lying metastable state that may be suitable for clock development.
In this work, we study relevant parameters of
Lu$^{2+}$, including transition matrix elements, lifetimes, polarizabilities, hyperfine constants and
the blackbody radiation (BBR) shift of the potential clock transition.

The properties of neutral, singly and doubly ionized lutetium were investigated recently in Refs.~\cite{dzuba-14,dzuba-14a,dzuba-14b,singapore-16,AB:16}.
\citet{dzuba-14} evaluated excitation
energies, ionization potentials and static dipole polarizabilities of lutetium  using a combination of the configuration interaction method and the all-order single-double coupled-cluster technique.
The calculations in \cite{dzuba-14} included Breit and quantum electrodynamic corrections.
\citet{dzuba-14a}  calculated excitation energies of the lowest states
of Lu$^+$ and Lu$^{2+}$  using a simplified version of the method used
in \cite{dzuba-14} but leading to  results of comparable accuracy.
\citet{dzuba-14b} studied a wide range of neutral atoms and ions
suitable for ultraprecise atomic optical clocks, in which the BBR shifts
of the clock transition frequencies are naturally suppressed,
and presented calculations of the BBR shift in Lu$^+$.
Atomic properties of Lu$^+$ were investigated by
\citet{singapore-16} where a joint experimental and theoretical investigation of Lu$^+$
as a clock candidate was reported and measurements relevant to practical clock operation were made. Calculations of scalar and tensor polarizabilities for clock states over a range of wavelengths were also given in \cite{singapore-16}.
\citet{AB:16} show that Lu$^{2+}$ is among the clock candidates with an upper $D$ state
for which shifts from rank 2 tensor interactions can be practically eliminated by operating at a judiciously chosen field-insensitive transition. 

Early publications include only few experimental \cite{1971,1979}
and theoretical \cite{migd-82,migd-92,biemont-99,biemont-04}
studies of Lu$^{2+}$.
The one-electron spectrum of Lu$^{2+}$ was
studied by \citeauthor{1971}. The $ns$
series was used to derive an ionization energy of
 169049$\pm$10~cm$^{-1}$. The ground state hyperfine constant was measured in the same work
 \cite{1971}. Subsequently, $5f-ng$ ($n$ = 5, 6) lines and $ng$ levels of Lu$^{2+}$ were reported by \citet{1979}.	
\citet{migd-82} reported calculations  of relativistic
model-potential ionization energies and transition probabilities
in the one-electron spectrum of doubly ionized lutetium employing
a technique that includes valence-core exchange and
correlation. The influence of polarization of the core by the
valence electron on ionization energies and transition
probabilities was also studied in \cite{migd-82}. Relativistic
ionization energies and fine structure intervals of $4f^{14}nl$
states in Lu$^{2+}$ ion were computed by \citet{migd-92}.
Weighted oscillator strengths and
radiative transition probabilities of Lu$^{2+}$ were evaluated by
\citet{biemont-99} using  relativistic Hartree-Fock (HFR) technique
described by Cowan (1981) in which
core-polarization effects are incorporated.
\citet{biemont-04} calculated Land\'{e} g-factor for over 1500
energy levels of doubly ionized lanthanides ($Z$ = 57-71) using the
HFR method combined with a least-squares fit of the
eigenvalues to observed energy levels; the resulting energies
were compared with previous experimental and theoretical
values.

In the present work, a relativistic all-order
method is used to calculate properties
of the first 30 excited $ns$, $np$,
$nd$, and $nf$ states of Lu$^{2+}$.
Electric-dipole matrix elements for
allowed transitions between low-lying $6s-np$, $6p-ns$, $6p-nd$, $5d-np$ and $5d-nf$  states of Lu$^{2+}$ are calculated and recommended values
are given for the electric-dipole matrix elements.
Line strengths, transition rates and lifetimes are calculated for the metastable $5d_{3/2}$ and $5d_{5/2}$ states.
Scalar and tensor polarizabilities of the
$6s$, $5d$ and $6p$ states are evaluated and recommended values of the polarizabilities are given together with uncertainties in these values.
Results for static scalar polarizabilities
of the $6s_{1/2}$ ground and $5d_{5/2}$ excited states are used
to determine the BBR shift of the $6s_{1/2}-5d_{5/2}$ clock transition frequency of Lu$^{2+}$.
Finally, we investigate the hyperfine structure in $^{175}$Lu$^{2+}$
and evaluate hyperfine  $A$ and $B$ constants for low-lying levels with principal quantum numbers $n \le 9$.

\section{Correlation energies of $\text{Lu}$~III}

Contributions to energies of Lu$^{2+}$ are listed in \tref{en-cor}.
These contributions include the zeroth-order Dirac-Fock energy $E^{(0)}$,
relativistic second- and third-order many-body perturbation
theory (MBPT) correlation energies $E^{(2)}$ and $E^{3)}$, all-order
energies in the single-double (SD) approximation $E^\text{SD}$ in which single and double
excitations of the Dirac-Fock wave function are summed to all
orders of perturbation theory and all-order single-double partial triple (SDpT) energies  $E^\text{SDpT}$
in which single, double and the dominant class of triple
excitations are summed to all orders. The SD energy
$E^\text{{SD}}$ includes $E^{(2)}$ completely but misses part of the third-order energy. The missing part $E^{(3)}_\text{{extra}}$ is included perturbatively
in the total SD energy.
The triple-excitations terms in the SDpT wave functions were
originally discussed by \citet{safr-cs,safr-alk}; these terms automatically
include the entire third-order energy and
substantially improve the accuracy of the SD functions.
The SDpT functions have been used in a number of publication
\cite{mar-05,mar-08,mar-11,mar-11a,SafSaf11,SafSaf12,SafSafCla12,SafSaf13,SafSaf14}
to evaluate multipole matrix elements.

As expected, the largest contribution to the correlation
energy comes from the second-order term $E^{(2)}$. This term is
relatively simple to calculate; thus, we calculate $E^{(2)}$ with
higher numerical accuracy than $E^{\text{SD}}$ and
$E^{\text{SDpT}}$ which are limited to partial waves with $l \le 6$.
The second-order energy $E^{(2)}$, by contrast, includes partial
waves up to $l_{\text{max}}=8$ and is extrapolated to account for
contributions from higher partial waves (see, for example,
\cite{SafSaf13,SafSaf14}). In the column headed $E^{(l>6)}$ of
\tref{en-cor}, we list the difference of the value of $E^{(2)}$
obtained with extrapolation and the value of $E^{(2)}$ evaluated with
$l_{\rm max}$ = 6.
The columns in \tref{en-cor} headed $B^{(1)}$ and $B^{(2)}$ list the
first-order Breit energy and the second-order Breit-Coulomb energy.

The third-order, SD and SDpT correlation energies are defined by
$E^{(3)}_\text{corr} = E^{(2)} + E^{(3)} + B^{(1)} + B^{(2)}$,
$E^\text{SD}_\text{corr} = E^\text{{SD}} +
E^{(3)}_\text{{extra}} + E^{(l>6)} + B^{(1)} + B^{(2)}$ and
$E^\text{SDpT}_\text{corr}$ = $E^\text{{SDpT}} + E^{(l>6)} + B^{(1)} + B^{(2)}$.
The relative correlation contributions $E_\text{corr}/(E^{(0)}+ E_\text{corr})$ in percent
are given in the  final three columns of \tref{en-cor}.
Comparing the values listed in the last three columns, we find that the largest
correlation contributions are for energies evaluated in the SDpT approximation.
The largest correlation contributions (about 7\%) are for the
$5d_{3/2}$ and the $5d_{5/2}$  levels. The smallest ones
(about 1.5\%) are for the $nf_{j}$ ($n$ 6 - 8) levels. For these
levels the Breit corrections  $B^{(1)}$ and $B^{(2)}$ are smaller than
for other
levels displayed in \tref{en-cor}.  We do not list QED correction,
however,  estimates of the QED corrections are included in totals given in
\tref{en-comp}.  The largest QED correction (30~cm$^{-1}$) is
for the $6s$ level.

\begin{table*}
\caption{\label{en-cor} Contributions to the energy levels of
Lu$^{+2}$ in cm$^{-1}$: $E^{(0)}$ is the zeroth-order Dirac-Fock
energy, $E^{(2)}$ and $E^{(3)}$ are the second-, and
third-order Coulomb correlation energies,
$E^\text{{SD}}$ and $E^{\text{SDpT}}$ are the Coulomb correlation energies in
the SD and SDpT approximations, $E^{(3)}_\text{{extra}}$ is the part of
the third-order energy missing in the SD approximation, $E^{(l>6)}$ is the
contribution to the second-order energy from partial waves with $l>6$ and
$B^{(1)}$  and $B^{(2)}$ are first-order Breit and second-order
Coulomb-Breit energies. The correlation energies are
$E^{(3)}_\text{corr} = E^{(2)} + E^{(3)} + B^{(1)} + B^{(2)}$,
$E^\text{SD}_\text{corr} =
E^\text{{SD}} + E^{(3)}_\text{{extra}} + E^{(l>6)} + B^{(1)} +
B^{(2)}$ and $E^\text{SDpT}_\text{corr} =  E^\text{{SDpT}} +
E^{(l>6)} + B^{(1)} + B^{(2)}$.
Relative contributions to the correlation
energy $E_\text{corr.}/(E^{(0)}+E_\text{corr.})$ in percent are given
in the last three columns.}
\begin{ruledtabular}\begin{tabular}{lrrrrrrrrrrrr}
\multicolumn{1}{c}{$nlj$ } & \multicolumn{1}{c}{$E^{(0)}$} &
\multicolumn{1}{c}{$E^{(2)}$} & \multicolumn{1}{c}{$E^{(3)}$} &
\multicolumn{1}{c}{$E^\text{{SD}}$} &
\multicolumn{1}{c}{$E^{(3)}_\text{{extra}}$} &
\multicolumn{1}{c}{$E^{\text{SDpT}}$} &
\multicolumn{1}{c}{$E^{(l>6)}$} & \multicolumn{1}{c}{$B^{(1)}$} &
\multicolumn{1}{c}{$B^{(2)}$} &
\multicolumn{1}{c}{$E^{(3)}_\text{corr}$} & \multicolumn{1}{c}{
$E^\text{SD}_\text{corr}$}&
\multicolumn{1}{c}{ $E^\text{SDpT}_\text{corr}$} \\
 \hline
 $ 6s_{1/2}$&   -160745&   -10006&   2041&  -9005&     340&   -8627&    -107&     209&   -354&     4.8&    5.2&    5.2\\[0.4pc]

 $ 5d_{3/2}$&   -152866&   -11922&    324& -10508&   -1378&   11288&    -395&     330&   -905&     7.4&    7.8&    7.4\\
 $ 5d_{5/2}$&   -150818&   -10714&    -53&  -9534&   -1434&   10341&    -371&     244&   -835&     7.0&    7.3&    7.0\\
 $ 6d_{3/2}$&    -74377&    -2585&    258&  -2307&    -173&   -2285&     -69&      71&   -173&     3.2&    3.4&    3.2\\
 $ 6d_{5/2}$&    -73679&    -2445&    204&  -2207&    -183&   -2285&     -67&      54&   -165&     3.1&    3.4&    3.2\\
 $ 7d_{3/2}$&    -45329&    -1150&    137&  -1042&     -64&   -1036&     -29&      32&    -75&     2.3&    2.5&    2.4\\
 $ 7d_{5/2}$&    -44994&    -1101&    116&  -1007&     -68&   -1036&     -29&      24&    -72&     2.2&    2.5&    2.4\\
 $ 8d_{3/2}$&    -30630&     -626&     81&   -580&      81&    -586&     -15&      17&    -40&     1.8&    1.7&    2.0\\
 $ 8d_{5/2}$&    -30442&     -603&     70&   -566&      70&    -572&     -15&      13&    -39&     1.8&    1.7&    2.0\\
 $ 9d_{3/2}$&    -22105&     -381&     52&   -360&      52&    -360&      -9&      10&    -24&     1.5&    1.5&    1.7\\
 $ 9d_{5/2}$&    -21990&     -368&     45&   -353&      45&    -357&      -9&       8&    -23&     1.5&    1.5&    1.7\\[0.4pc]

 $ 7s_{1/2}$&    -79912&    -2972&    608&  -2596&      85&   -2510&     -33&      70&   -114&     2.9&    3.1&    3.1\\
 $ 8s_{1/2}$&    -48159&    -1318&    270&  -1139&      35&   -1105&     -15&      33&    -52&     2.2&    2.3&    2.3\\
 $ 9s_{1/2}$&    -32240&     -705&    145&   -608&      18&    -590&      -8&      18&    -28&     1.7&    1.9&    1.9\\[0.4pc]

 $ 6p_{1/2}$&   -125045&    -6161&   1024&  -5947&     170&   -5720&     -60&     179&   -199&     4.0&    4.5&    4.4\\
 $ 6p_{3/2}$&   -119552&    -5171&    767&  -5024&      87&   -4873&     -53&     123&   -181&     3.6&    4.1&    4.0\\
 $ 7p_{1/2}$&    -66610&    -2169&    343&  -2224&      47&   -2144&     -22&      70&    -77&     2.7&    3.2&    3.2\\
 $ 7p_{3/2}$&    -64459&    -1884&    269&  -1929&      22&   -1871&     -21&      49&    -72&     2.5&    2.9&    2.9\\
 $ 8p_{1/2}$&    -41708&    -1046&    160&   -985&      22&    -954&     -11&      35&    -38&     2.1&    2.3&    2.3\\
 $ 8p_{3/2}$&    -40635&     -923&    128&   -622&     128&    -618&     -10&      25&    -36&     1.9&    1.3&    1.5\\
 $ 9p_{1/2}$&    -28623&     -588&     88&   -439&      12&    -434&      -6&      20&    -22&     1.7&    1.4&    1.4\\
 $ 9p_{3/2}$&    -28010&     -524&     71&   -403&       6&    -402&      -6&      14&    -21&     1.5&    1.4&    1.4\\[0.4pc]

 $ 5f_{5/2}$&    -61961&    -1546&    258&  -1880&     258&   -1694&     -17&       6&    -24&     2.1&    2.6&    2.7\\
 $ 5f_{7/2}$&    -61949&    -1516&    204&  -1833&     204&   -1663&     -17&       5&    -27&     2.1&    2.6&    2.7\\
 $ 6f_{5/2}$&    -39685&     -860&    132&   -589&     132&    -552&     -12&       4&    -18&     1.8&    1.2&    1.4\\
 $ 6f_{7/2}$&    -39676&     -838&    112&   -388&     112&    -370&     -12&       4&    -19&     1.8&    0.8&    1.0\\
 $ 7f_{5/2}$&    -27558&     -472&     67&   -468&      36&    -468&      -8&       3&    -12&     1.5&    1.6&    1.7\\
 $ 7f_{7/2}$&    -27551&     -457&     54&   -435&      31&    -435&      -8&       2&    -13&     1.5&    1.5&    1.6\\
 $ 8f_{5/2}$&    -20241&     -302&     38&   -302&      22&    -302&      -5&       2&     -8&     1.3&    1.4&    1.5\\
 $ 8f_{7/2}$&    -20236&     -292&     30&   -282&      19&    -282&      -5&       2&     -9&     1.3&    1.3&    1.4\\
 \end{tabular}
\end{ruledtabular}
\end{table*}

Recommended energies of states of Lu$^{2+}$ from the National
Institute of Standards and Technology (NIST) database
\cite{nist-web} are given in the column of \tref{en-comp}
headed $E_\text{NIST}$. This column is followed by theoretical removal energies
$E^{(3)}_\text{tot} = E^{(0)} +
E^{(3)}_\text{corr} + E^\text{(QED)}$,
$E^\text{SD}_\text{tot} = E^{(0)} + E^\text{SD}_\text{corr} +
E^\text{(QED)}$  and $E^\text{SDpT}_\text{tot} =
E^{(0)} + E^\text{SDpT}_\text{corr}+ E^\text{(QED)}$.
Relative differences (in percent) between the theoretical third-order and
all-order energies and the experimental data,
$\delta E = (E_{\rm tot}-E_\text{NIST})/E_\text{NIST}$, are given in
the final three columns of \tref{en-comp}. The smallest
differences are obtained using the SDpT method.
Results for the $8d$, $9d$, $8p$, and $9p$ levels
are not included in \tref{en-comp} since these
levels are not included in the NIST database \cite{nist-web}.

 \begin{table}
\caption{\label{en-comp} The total removal  energies (cm$^{-1}$)
of  Lu$^{+2}$ ($E^{(3)}_\text{ tot} = E^{(0)} + E^{(3)}_\text{corr} +
E^\text{(QED)}$, $E^\text{SD}_\text{tot} =
E^{(0)} +  E^\text{SD}_\text{corr} + E^\text{(QED)}$
and $E^\text{SDpT}_\text{tot} = E^{(0)} + E^\text{SDpT}_\text{corr}  +
E^\text{(QED)}$) are compared with recommended NIST
energies $E_\text{{NIST}}$ \cite{nist-web}.
The relative difference $\delta E = (E_\text{tot} - E_\text{{NIST}})/E_\text{{NIST}}$
in percent is given in the three last columns.}
\begin{ruledtabular}
\begin{tabular}{lrrrrrrr}
\multicolumn{1}{c}{$nlj$ } & \multicolumn{1}{c}{$E_\text{{NIST}}$}
& \multicolumn{1}{c}{$E^{(3)}_\text{ tot}$} &
\multicolumn{1}{c}{$E^\text{{SD}}_\text{tot}$} &
\multicolumn{1}{c}{$E^\text{{SDpT}}_\text{tot}$} &
\multicolumn{1}{c}{$\delta E^{(3)}$} & \multicolumn{1}{c}{$\delta
E^\text{{SD}}$}&
\multicolumn{1}{c}{$\delta E^\text{{SDpT}}$} \\
\hline
 $ 6s_{1/2}$&   -169014&   -168817&  -169625&  -169587&      -0.12    &  0.36&    0.34\\[0.4pc]
 $ 5d_{3/2}$&   -163306&   -165040&  -165723&  -165125&       1.05    &  1.46&    1.10\\
 $ 5d_{5/2}$&   -160366&   -162177&  -162748&  -162122&       1.12    &  1.46&    1.08\\
 $ 6d_{3/2}$&    -76692&    -76806&   -77028&   -76832&       0.15    &  0.44&    0.18\\
 $ 6d_{5/2}$&    -75906&    -76031&   -76247&   -76141&       0.16    &  0.45&    0.31\\
 $ 7d_{3/2}$&    -46392&    -46385&   -46507&   -46437&      -0.01    &  0.25&    0.10\\
 $ 7d_{5/2}$&    -46033&    -46026&   -46145&   -46105&      -0.01    &  0.24&    0.16\\[0.4pc]
 $ 7s_{1/2}$&    -82333&    -82313&   -82493&   -82493&      -0.02    &  0.19&    0.19\\
 $ 8s_{1/2}$&    -49229&    -49225&   -49297&   -49298&      -0.01    &  0.14&    0.14\\
 $ 9s_{1/2}$&    -32804&    -32810&   -32848&   -32848&       0.02    &  0.13&    0.13\\[0.4pc]
 $ 6p_{1/2}$&   -130613&   -130202&  -130902&  -130844&      -0.32    &  0.22&    0.18\\
 $ 6p_{3/2}$&   -124309&   -124014&  -124599&  -124536&      -0.24    &  0.23&    0.18\\
 $ 7p_{1/2}$&    -68657&    -68443&   -68816&   -68783&      -0.31    &  0.23&    0.18\\
 $ 7p_{3/2}$&    -66203&    -66097&   -66409&   -66374&      -0.16    &  0.31&    0.26\\[0.4pc]

 $ 5f_{5/2}$&    -63423&    -63268&   -63619&   -63691&      -0.25    &  0.31&    0.42\\
 $ 5f_{7/2}$&    -63310&    -63283&   -63617&   -63651&      -0.04    &  0.48&    0.54\\
 $ 6f_{5/2}$&    -40214&    -40427&   -40168&   -40262&       0.53    & -0.12&    0.12\\
 $ 6f_{7/2}$&    -39961&    -40418&   -39980&   -40074&       1.13    &  0.05&    0.28\\
 $ 7f_{5/2}$&    -27944&    -27971&   -28006&   -28042&       0.10    &  0.22&    0.35\\
 $ 7f_{7/2}$&    -27922&    -27964&   -27973&   -28004&       0.15    &  0.18&    0.29\\
 $ 8f_{5/2}$&    -20500&    -20511&   -20532&   -20554&       0.05    &  0.15&    0.26\\
 $ 8f_{7/2}$&    -20484&    -20506&   -20512&   -20530&       0.11    &  0.14&    0.23\\
 \end{tabular}
\end{ruledtabular}
\end{table}

\section{Electric-dipole matrix elements and lifetimes}

In \tref{tab-dip}, recommended values of  reduced electric dipole matrix elements of 49 $6s-np$,  $6p-ns$, $6p-nd$, $5d-np$ and $5d-nf$ transitions are presented. The absolute values in atomic units ($e\, a_0$) are given in all cases.
Matrix elements for eight $6s-np$ ($n = 6-9$) transitions, eight $6p-ns$ ($n = 6-9$) transitions, nine
$6p-nd$ ($n = 5-7$) transitions, twelve $5d-np$ ($n = 6-9$) transitions and twelve $5d-nf$ ($n = 5-8$) transitions are evaluated. The recommended values of the
matrix elements for these transitions are listed in the column headed ``Final".
To determine these values and estimate the corresponding uncertainties, we carried out a series of calculations using different methods of
increasing accuracy: lowest-order DF, second-order MBPT, third-order MBPT and
all-order methods. The MBPT calculations were carried out using the
method described in Ref.~\cite{adndt-96}.  Comparisons of
values obtained in different approximations allow us to evaluate
the size of the second, third and higher-order correlation
corrections as well as estimate uncertainties in the final
values.
The evaluation of uncertainty of the matrix elements using this approach was
described in detail in Refs.~\cite{SafSaf11rb,SafSaf12}. The uncertainty evaluation is based on four different
all-order calculations. These  include two \textit{ab initio} all-order calculations, carried out  with and without the partial triple excitations and two
calculations that included semiempirical estimates of high-order
correlation corrections starting from both \textit{ab initio}
runs. We use the differences in these four values
to  estimate the uncertainty in the final results for each transition. The estimates are based on an algorithm that accounted for the dominant contributions.

The column labeled ``Rel.unc'' in \tref{tab-dip} gives relative uncertainties of the final reduced matrix elements in percent. We could not carry out the scaling for the $8d$, $9d$, $8p$, and $9p$ levels since there are no experimental energy values available.
As a result, we used only the SD
and SDpT values to determine the uncertainties listed in two
last columns of \tref{tab-dip}. Uncertainties given in
the column ``Rel.unc'' of \tref{tab-dip} are in the range 0.38\% - 2\%.
However, there are  matrix elements with uncertainties equal to
19\% and 27\%. Such large uncertainties occur only for very
small matrix elements, 0.0225 and 0.0099, respectively, where the relative contribution
of correlation corrections is very large. Uncertainties for larger matrix elements,
such as the $6s - 6p$ matrix element, are much smaller (0.38\% and  0.40\%).
Uncertainties in matrix elements of $5d-nf$ transitions are smaller than those of the $5d-np$ transitions.

Line strengths $S$~(a.u.), transition rates $A$~(s$^{-1}$) and lifetimes $\tau$~(s) of the metastable $5d$ states of the Lu$^{2+}$ ion, evaluated by
the scaled SD method, are listed in Table~\ref{lif}.

\begin{table*}
\caption{\label{tab-dip}Recommended values of reduced
electric-dipole matrix elements in atomic units. The lowest-order DF,
all-order SD and SDpT values are listed; the label ``sc''
indicates the scaled values. Recommended values of the matrix elements
given in the column headed ``Final" and the corresponding uncertainties
are listed under the heading ``Unc.".  The last
column gives relative uncertainties ``Rel. unc."  in
percent. }
\begin{ruledtabular}
\begin{tabular}{llrrrrrrcc}
\multicolumn{2}{c}{Transition}&
\multicolumn{1}{c}{DHF}&
\multicolumn{1}{c}{SD}&
\multicolumn{1}{c}{SDsc}&
\multicolumn{1}{c}{SDpT}&
\multicolumn{1}{c}{SDpTsc}&
\multicolumn{1}{c}{Final}&
\multicolumn{1}{c}{Unc.}&
\multicolumn{1}{c}{Rel. unc.}\\
\hline
 $6s_{1/2}$&$  6p_{1/2}$&   2.7279  &  2.3177   & 2.3264  &  2.3193 &   2.3236  &  2.3177  &  0.0087  &  0.38\\
 $6s_{1/2}$&$  7p_{1/2}$&   0.0107  &  0.1716   & 0.1703  &  0.1716 &   0.1702  &  0.1716  &  0.0014  &  0.82\\
 $6s_{1/2}$&$  8p_{1/2}$&   0.0339  &  0.1343   &         &  0.1359 &           &  0.1343  &  0.0016  &  1.20\\
 $6s_{1/2}$&$  9p_{1/2}$&   0.0289  &  0.0398   &         &  0.0409 &           &  0.0398  &  0.0011  &  2.76\\[0.3pc]

 $6s_{1/2}$&$  6p_{3/2}$&    3.8241 &   3.2695  &  3.2826 &   3.2721&    3.2786 &   3.2695 &   0.0131 &  0.40\\
 $6s_{1/2}$&$  7p_{3/2}$&    0.2041 &   0.0225  &  0.0186 &   0.0237&    0.0182 &   0.0225 &   0.0043 &  19.1\\
 $6s_{1/2}$&$  8p_{3/2}$&    0.0658 &   0.0803  &         &   0.0838&           &   0.0838 &   0.0035 &  4.15\\
 $6s_{1/2}$&$  9p_{3/2}$&    0.0333 &   0.0099  &         &   0.0072&           &   0.0099 &   0.0027 &  27.3\\ [0.3pc]

$ 6p_{1/2}$&$   6s_{1/2}$&     2.7279&    2.3177&    2.3260&    2.3193&    2.3236&    2.3177&    0.0083&    0.36 \\
$ 6p_{1/2}$&$   7s_{1/2}$&     1.6729&    1.6315&    1.6265&    1.6240&    1.6246&    1.6265&    0.0071&    0.44 \\
$ 6p_{1/2}$&$   8s_{1/2}$&     0.4753&    0.4807&    0.4803&    0.4721&    0.4723&    0.4807&    0.0117&    2.43 \\
$ 6p_{1/2}$&$   9s_{1/2}$&     0.2595&    0.2678&    0.2677&    0.2590&    0.2594&    0.2678&    0.0088&    3.29 \\[0.3pc]

$ 6p_{1/2}$&$   5d_{3/2}$&     2.4588&    2.0696&    2.0701&    2.0459&    2.0598&    2.0696&    0.0343&    1.66 \\
$ 6p_{1/2}$&$   6d_{3/2}$&     3.6470&    3.3671&    3.3753&    3.3876&    3.3975&    3.3671&    0.0313&    0.93 \\
$ 6p_{1/2}$&$   7d_{3/2}$&     1.0403&    0.9077&    0.9049&    0.9072&    0.9040&    0.9077&    0.0032&    0.35 \\[0.3pc]

$ 6p_{3/2}$&$   6s_{1/2}$&     3.8241&    3.2695&    3.2827&    3.2721&    3.2786&    3.2695&    0.0132&    0.40 \\
$ 6p_{3/2}$&$   7s_{1/2}$&     2.8163&    2.7497&    2.7413&    2.7083&    2.7081&    2.7413&    0.0469&    1.71 \\
$ 6p_{3/2}$&$   8s_{1/2}$&     0.7216&    0.7151&    0.7153&    0.6860&    0.6865&    0.7153&    0.0415&    5.80 \\
$ 6p_{3/2}$&$   9s_{1/2}$&     0.3857&    0.3838&    0.3843&    0.3618&    0.3625&    0.3838&    0.0318&    8.29 \\ [0.3pc]

$ 6p_{3/2}$&$   5d_{3/2}$&     1.0423&    0.8898&    0.8901&    0.8788&    0.8854&    0.8898&    0.0160&    1.80 \\
$ 6p_{3/2}$&$   6d_{3/2}$&     1.8135&    1.6845&    1.6876&    1.6921&    1.6958&    1.6845&    0.0113&    0.67 \\
$ 6p_{3/2}$&$   7d_{3/2}$&     0.4490&    0.3845&    0.3828&    0.3835&    0.3815&    0.3845&    0.0030&    0.78 \\ [0.3pc]

$ 6p_{3/2}$&$   5d_{5/2}$&     3.2482&    2.7973&    2.7962&    2.7632&    2.7826&    2.7973&    0.0467&    1.67 \\
$ 6p_{3/2}$&$   6d_{5/2}$&     5.3437&    4.9661&    4.9766&    4.9900&    4.9928&    4.9661&    0.0228&    0.46 \\
$ 6p_{3/2}$&$   7d_{5/2}$&     1.3809&    1.1953&    1.1892&    1.1922&    1.1896&    1.1953&    0.0057&    0.48 \\ [0.3pc]

 $5d_{3/2}$&$  6p_{1/2}$&   2.4588  &  2.0696   & 2.0694  &  2.0459 &   2.0598  &  2.0696  &  0.0332  &  1.60  \\
 $5d_{3/2}$&$  7p_{1/2}$&   0.2751  &  0.1094   & 0.1174  &  0.1159 &   0.1133  &  0.1094  &  0.0080  &  7.31  \\
 $5d_{3/2}$&$  8p_{1/2}$&   0.1487  &  0.0905   &         &  0.0941 &           &  0.0905  &  0.0036  &  3.98  \\
 $5d_{3/2}$&$  9p_{1/2}$&   0.1006  &  0.0709   &         &  0.0729 &           &  0.0709  &  0.0020  &  2.82  \\[0.3pc]

 $5d_{3/2}$&$  6p_{3/2}$&    1.0423 &   0.8898  &  0.8897 &   0.8788&    0.8854 &   0.8898 &   0.0155 &   1.74 \\
 $5d_{3/2}$&$  7p_{3/2}$&    0.1530 &   0.0924  &  0.0959 &   0.0951&    0.0951 &   0.0924 &   0.0035 &   3.79 \\
 $5d_{3/2}$&$  8p_{3/2}$&    0.0819 &   0.0728  &         &   0.0735&           &   0.0728 &   0.0007 &   0.96 \\
 $5d_{3/2}$&$  9p_{3/2}$&    0.0550 &   0.0420  &         &   0.0425&           &   0.0420 &   0.0005 &   1.19 \\[0.3pc]

 $5d_{5/2}$&$  6p_{3/2}$&    3.2482 &   2.7973  &  2.7953 &   2.7632&    2.7826 &   2.7973 &   0.0454 &   1.62 \\
 $5d_{5/2}$&$  7p_{3/2}$&    0.4451 &   0.2694  &  0.2814 &   0.2794&    0.2795 &   0.2694 &   0.0120 &   4.45 \\
 $5d_{5/2}$&$  8p_{3/2}$&    0.2382 &   0.2246  &         &   0.2300&           &   0.2246 &   0.0054 &   2.40 \\
 $5d_{5/2}$&$  9p_{3/2}$&    0.1601 &   0.1297  &         &   0.1317&           &   0.1297 &   0.0020 &   1.54 \\[0.3pc]

 $5d_{3/2}$&$  5f_{5/2}$&   2.5512  &  2.0403   & 2.0057  &  1.9972 &   1.9979  &  2.0403  &  0.0431  &  2.11  \\
 $5d_{3/2}$&$  6f_{5/2}$&   1.3900  &  1.2302   & 1.2303  &  1.2221 &   1.2274  &  1.2302  &  0.0117  &  0.95  \\
 $5d_{3/2}$&$  7f_{5/2}$&   0.9226  &  0.7511   & 0.7529  &  0.7464 &   0.7494  &  0.7511  &  0.0065  &  0.86  \\
 $5d_{3/2}$&$  8f_{5/2}$&   0.6769  &  0.5406   & 0.5427  &  0.5361 &   0.5380  &  0.5406  &  0.0066  &  1.22  \\[0.3pc]

 $5d_{5/2}$&$  5f_{5/2}$&   0.7099  &  0.5853   & 0.5754  &  0.5733 &   0.5734  &  0.5853  &  0.0099  &  1.69  \\
 $5d_{5/2}$&$  6f_{5/2}$&   0.3816  &  0.3369   & 0.3368  &  0.3348 &   0.3361  &  0.3369  &  0.0029  &  0.86  \\
 $5d_{5/2}$&$  7f_{5/2}$&   0.2518  &  0.2021   & 0.2027  &  0.2010 &   0.2018  &  0.2021  &  0.0017  &  0.84  \\
 $5d_{5/2}$&$  8f_{5/2}$&   0.1842  &  0.1441   & 0.1447  &  0.1430 &   0.1435  &  0.1441  &  0.0017  &  1.18\\[0.3pc]

 $5d_{5/2}$&$  5f_{7/2}$&    3.1693 &   2.5743  &  2.5256 &   2.5224&    2.5161 &   2.5743 &   0.0582 &   2.26 \\
 $5d_{5/2}$&$  6f_{7/2}$&    1.7028 &   1.5869  &  1.5889 &   1.5806&    1.5876 &   1.5889 &   0.0083 &   0.52 \\
 $5d_{5/2}$&$  7f_{7/2}$&    1.1235 &   0.9305  &  0.9326 &   0.9253&    0.9287 &   0.9305 &   0.0073 &   0.78 \\
 $5d_{5/2}$&$  8f_{7/2}$&    0.8218 &   0.6669  &  0.6693 &   0.6644&    0.6665 &   0.6669 &   0.0049 &   0.73 \\
\end{tabular}
\end{ruledtabular}
\end{table*}

\begin{table}
\caption{Multipolarities (MP), wavelengths $\lambda$~(\AA), line strengths $S$~(a.u.),
transition rates $A$~(s$^{-1}$) and lifetimes $\tau$~(s) are given for
transitions between metastable $5d_{3/2}$ and $5d_{5/2}$ states and the $6s_{1/2}$ ground state of Lu$^{2+}$. Numbers in square brackets represent
powers of 10. \label{lif}}
\begin{ruledtabular}
\begin{tabular}{lcccccc}
\multicolumn{1}{c}{Transition}
      & MP & $\lambda$~(\AA) & S~(a.u.)& A~(s$^{-1}$)  & $\tau$~(s) \\
\hline
 $5d_{3/2} -  6s_{1/2}$ & E2 & 17520   & 37.831   & 6.415[-3]& 155.9 \\
 $5d_{3/2} -  6s_{1/2}$ & M1 & 17520   & 1.39[-7] & 1.744[-7]& \\
 $5d_{5/2} -  5d_{3/2}$ & M1 & 34011   & 2.4013   & 2.744[-1]& 3.029 \\
 $5d_{5/2} -  6s_{1/2}$ & E2 & 11564   & 61.234   & 5.527[-2]& \\
 $5d_{5/2} -  5d_{3/2}$ & E2 & 34011   & 12.016   & 4.928[-4]& \\
 \end{tabular}
 \end{ruledtabular}
 \end{table}

%
%\begin{table*}
%\caption{\label{lif} The lifetimes $\tau$ (sec) of the $5d$ states.
%Corresponding energies  (cm$^{-1}$), wavelengths
%($\lambda$ in \AA),  transition rates $A_r$ (s$^{-1}$) and  branching
%ratios  are also listed.
%The numbers in brackets represent powers of 10.}
%\begin{ruledtabular}
%\begin{tabular}{lllrrrrrrrrr}
%\multicolumn{1}{c}{Level}& \multicolumn{2}{c}{Transition}&
%\multicolumn{1}{c}{Energy}& \multicolumn{1}{c}{Energy}&
%\multicolumn{2}{c}{Trans. energy}& \multicolumn{1}{c}{$\lambda$}&
%\multicolumn{1}{c}{$Z^{\rm CI+all}$}&
%\multicolumn{1}{c}{$A_{r}^{\rm  CI+all}$}& \multicolumn{1}{c}{Br.
%ratio}&
%\multicolumn{1}{c}{$\tau^{\rm  CI+all}$}\\
%\multicolumn{3}{c}{}&
%\multicolumn{1}{c}{cm$^{-1}$}&
%\multicolumn{1}{c}{cm$^{-1}$}&
%\multicolumn{1}{c}{cm$^{-1}$}&
%\multicolumn{1}{c}{}&
%\multicolumn{1}{c}{\AA}&
%\multicolumn{1}{c}{a.u.}&
%\multicolumn{1}{c}{s$^{-1}$}&
%\multicolumn{1}{c}{}&
%\multicolumn{1}{c}{}\\
%\hline
% $5d_{3/2} $&  $6s_{1/2} $&    $5d\ ^2D_{3/2} $&    0.0  &    5708   &    5708  &  E2     &   17520&    6.15070&   6.416[-3]& 6.416[-3] &155.9~sec   \\
%  $5d_{3/2} $&  $6s_{1/2} $&    $5d\ ^2D_{3/2} $&    0.0  &    5708   &    5708  &  M1     &    17520&    0.00037&  1.753[-7]&           & \\[0.4pc]
%
%  $5d_{5/2} $&  $6s_{1/2} $&    $5d\ ^2D_{5/2} $&    0.0  &    8648    &   8648  &  E2     &    11564&     7.82520&  5.528[-2]  & 5.528[-2]& 18.09~sec\\
%  $5d_{5/2} $&  $5d_{3/2} $&    $5d\ ^2D_{5/2} $&   5708  &   8648    &   2940  &  E2     &     34011&     3.46640&   4.928[-5] &         &            \\
%\end{tabular}
%\end{ruledtabular}
%\end{table*}
\begin{table}
\caption{\label{tab-6s} Contributions to scalar
polarizabilities of the $6s$ state of Lu$^{2+}$ in units $a_0^3$. The
dominant contributions are listed
separately with the corresponding absolute values of
electric-dipole reduced matrix elements given in columns labeled
$D$. The  experimental \cite{nist-web} transition energies are
given in columns $\Delta E$.
 Uncertainties are given in parenthesis. }
\begin{ruledtabular}
\begin{tabular}{lrrr}
 \multicolumn{1}{c}{Contr.} &
 \multicolumn{1}{c}{$\Delta E$} &
 \multicolumn{1}{c}{$D$} &
 \multicolumn{1}{c}{$\alpha^{E1}_{0}(6s)$}\\
 [0.3pc]\hline
   $6p_{1/2}$&  38400.61  &   2.3177 &        10.234(78) \\
   $7p_{1/2}$&  100357.09 &   0.1716 &         0.021(0)     \\
   $(8-26)p_{1/2}$ & &&    0.014(0)                        \\[0.4pc]

    $6p_{3/2}$&   44705.21 &   3.2695 &          17.493(140)\\
    $7p_{3/2}$&  102810.82 &  -0.0225 &          0.00(0)    \\
    $(8-26)p_{3/2}$&&&     0.034(0)     \\[0.4pc]
   CORE      &     &&      4.265 (080)  \\
   Term-vc   &     &&      -0.678(0)   \\
   Tail      &     &&       0.00       \\[0.4pc]
  Total      &     &&       31.91(18)\\
\end{tabular}
\end{ruledtabular}
\end{table}

%%%%%%%%%%%%%%%%%%%%%%%%%%%%%%%%%%%%%%%%%%%%%%%%%%%%%%%%%%%%%%%%%%%%%%%%%%%%%%%%%%%%%%%%%%%%%%%%%%%%%%%%%%%%%%%%%%%

\begin{table*}
\caption{\label{tab-5d} Contributions to the scalar and tensor
polarizabilities of the $5d_{3/2}$ and $5d_{5/2}$ state of
Lu$^{2+}$ in $a_0^3$. Uncertainties are given in parenthesis.}
\begin{ruledtabular}
\begin{tabular}{lrrlrr}
 \multicolumn{1}{c}{Contr.} &
 \multicolumn{1}{c}{$\alpha_{0}(5d_{3/2})$}&
  \multicolumn{1}{c}{$\alpha_{2}(5d_{3/2})$}&
 \multicolumn{1}{c}{Contr.}&
\multicolumn{1}{c}{$\alpha_{0}(5d_{5/2})$}&
 \multicolumn{1}{c}{$\alpha_{2}(5d_{5/2})$}\\
  [0.3pc]\hline
 $6p_{1/2}$&       4.792(153)&  -4.792(153)  &   $6p_{3/2}$      &  5.292(171)&  -5.292(171) \\
 $7p_{1/2}$&       0.005(1)  &  -0.005(1)    &   $7p_{3/2}$      &  0.019(2)  &  -0.019(2)   \\
 $(8-26)p_{1/2}$&  0.003(0)  &  -0.003(0)    &   $(9-26)p_{3/2}$ &   0.015(0) &   -0.015(0) \\[0.4pc]

 $6p_{3/2}$&       0.743(26) &  0.594(21)    &   $5f_{5/2}$      & 0.086(3)   & 0.098(3)  \\
 $7p_{3/2}$&       0.003(0)  &  0.003(0)     &   $6f_{5/2}$      & 0.023(0)   & 0.026(0)  \\
 $(9-26)p_{3/2}$&  0.000(0)  &  0.001(0)     &   $7f_{5/2}$      & 0.009(0)   & 0.008(0)  \\
                &            &               &   $8f_{5/2}$      & 0.089(1)   & 0.078(1)  \\
                &            &               &   $(9-26)f_{5/2}$ & 0.002(0)   & 0.005(0)   \\[0.4pc]

$5f_{5/2}$&        1.525(64) & -0.305(13)   &   $5f_{7/2}$       & 1.665(75)  & -0.595(27)  \\
$6f_{5/2}$&        0.450(9)  & -0.090(2)    &   $6f_{7/2}$       & 0.511(5)   & -0.183(2)  \\
$7f_{5/2}$&        0.152(3)  & -0.030(1)    &   $7f_{7/2}$       & 0.159(2)   & -0.057(1)  \\
$8f_{5/2}$&        0.075(2)  & -0.015(0)    &   $8f_{7/2}$       & 0.078(1)   & -0.028(0)  \\
$9f_{5/2}$&        0.043(0)  & -0.009(0)    &   $9f_{7/2}$       & 0.044(0)   & -0.016(0)  \\
$(10-26)f_{5/2}$&   0.181(0) &  -0.037(0)   &   $(10-26)f_{7/2}$ & 0.169(0)   &  -0.048(0)\\[0.4pc]
  CORE     &       4.264(080)& 0.0          &   CORE             & 4.264(080) &  0.0   \\
  Term-vc  &      -0.261     & 0.0          &   Term-vc          & -0.355(0)  &  0.0   \\
  Tail     &      0.0        & 0.0          &   Tail             &  0.0       &  0.0   \\[0.4pc]
  Total    &    11.98(19)    & -4.69(15)    &    Total           &  12.09(20) &  -6.03(17)   \\
\end{tabular}
\end{ruledtabular}
\end{table*}

\begin{table*}
\caption{\label{tab-6p} Contributions to the scalar and tensor
polarizabilities of the $6p_{1/2}$ and $6p_{3/2}$ state of
Lu$^{2+}$ in $a_0^3$. Uncertainties are given in parenthesis.}
\begin{ruledtabular}
\begin{tabular}{lrlrr}
 \multicolumn{1}{c}{Contr.} &
 \multicolumn{1}{c}{$\alpha_{0}(6p_{1/2})$}&
 \multicolumn{1}{c}{Contr.}&
\multicolumn{1}{c}{$\alpha_{0}(6p_{3/2})$}&
 \multicolumn{1}{c}{$\alpha_{2}(6p_{3/2})$}\\
  [0.3pc]\hline
 $6s_{1/2}$     &-10.234(74)&  $6s_{1/2}$     & -8.747(70) & 8.747(70)   \\
 $7s_{1/2}$     &  4.009(35)&  $7s_{1/2}$     &  6.549(224)& -6.549(224)  \\
 $(8-26)s_{1/2}$&  0.344(10)&  $(8-26)s_{1/2}$&   0.389(31) & -0.389(31)    \\[0.4pc]

               &              & $5d_{3/2}$      &-0.743(27)   & -0.594(21)  \\
               &              & $6d_{3/2}$      & 2.180(29)   & 1.744(23)  \\
               &              & $7d_{3/2}$      & 0.069(1)    & 0.056(1)  \\
               &              & $8d_{3/2}$      & 0.013(0)    & 0.010(0)  \\
               &              & $(9-26)d_{3/2}$ & 0.021(0)    & 0.017(0)   \\[0.4pc]

$5d_{3/2}$&       -9.585(318)&  $5d_{5/2}$       &-7.938(265) &  1.588(53)  \\
$6d_{3/2}$&        15.382(286)  &  $6d_{5/2}$    & 18.638(171)& -3.728(34)  \\
$7d_{3/2}$&        0.716(5)  &  $7d_{5/2}$       & 0.668(6)   & -0.134(1)  \\
$8d_{3/2}$&        0.154(0)  &  $8d_{5/2}$       & 0.131(0)   & -0.026(0)  \\
$9d_{3/2}$&        0.055(0)  &  $9d_{5/2}$       & 0.046(0)   & -0.009(0)  \\
$(10-26)d_{3/2}$&   0.181(0) &  $(10-26)d_{5/2}$ & 0.096(0)   &  -0.019(0)\\[0.4pc]
  CORE     &       4.264(080)&  CORE             & 4.264(080) &  0.0   \\
  Term-vc  &      -0.001(0)  &  Term-vc          & 0.0(0)     &  0.0   \\
  Tail     &      0.0        &  Tail             &  0.0       &  0.0   \\[0.4pc]
  Total    &    5.24(44)     &   Total           &  15.63(41) &  0.71(25)   \\
\end{tabular}
\end{ruledtabular}
\end{table*}

\section{Scalar and tensor excited state polarizabilities}

The scalar $\alpha_{0}(v)$  and tensor $\alpha_{2}(v) $
polarizability of an excited state with one valence electron $v$ are given by
\begin{equation}
\alpha _{\text{0}}(v)\
=\frac{2}{3(2j_{v}+1)}\sum_{nlj}\frac{|\langle
v||rC_{1}||nlj\rangle |^{2}}{E_{nlj}-E_{v}} \label{one}
\end{equation}
and
\begin{align}
\alpha _{2}(v) & \ =(-1)^{j_{v}}\sqrt{\frac{40j_{v}(2j_{v}-1)}{%
3(j_{v}+1)(2j_{v}+1)(2j_{v}+3)}}\   \nonumber\\
& \times \sum_{nlj}(-1)^{j}\left\{
\begin{array}{lll}
j_{v} & 1 & j \\
1 & j_{v} & 2
\end{array}
\right\} \frac{|\langle v||rC_{1}||nlj\rangle |^{2}}{E_{nlj}-E_{v}} , \label{two}
\end{align}
where  $C_{kq}(\hat{r})$ is a normalized spherical harmonic and
where the indices $nlj$ in the sums range over $np_{j}$, $nd_{j}$, and $nf_j$.

The polarizabilities in Eqs.~(\ref{one},\ref{two}) can be separated into two parts:
a dominant term from intermediate valence-excited states and
a contribution from core-excited states.
The second term is smaller than the former by
several orders of magnitude and is evaluated here in the
random-phase approximation \cite{RPA}. The dominant valence
contribution is calculated using the sum-over-state approach using our theoretical
recommended values of the matrix elements and  energies  from the
NIST database \cite{nist-web}. Uncertainties in the polarizability
contributions are obtained from the uncertainties in the
corresponding matrix elements.

Contributions to scalar
polarizabilities of the $6s_{1/2}$ state of Lu$^{2+}$ are given in
\tref{tab-6s}. The corresponding uncertainty is given in parenthesis.
The $6s_{1/2}- 6p_{1/2}$  and $6s_{1/2}- 6p_{3/2}$ transitions account
for  99.7\% of the final value of $\alpha_{0}(6s_{1/2})$.
%%%%%%%%%%%%%%%%%%%%%%%%%%%%%%%%%%%%%

Contributions to scalar and tensor polarizabilities of the $5d_{3/2}$ and
$5d_{5/2}$ levels
 of Lu$^{2+}$ are given in \tref{tab-5d}.
The largest contribution  to $\alpha_{0}(5d_{3/2})$  (88.5\%)
is from the three transitions $5d_{3/2} - 6p_{1/2}$,
$5d_{3/2} - 6p_{3/2}$ and $5d_{3/2} - 5f_{5/2}$. Similarly,
the largest contribution to $\alpha_{0}(5d_{5/2})$  (86.3\%)
is from the three transitions $5d_{5/2} - 6p_{3/2}$,
$5d_{5/2} - 5f_{5/2}$ and $5d_{5/2} - 5f_{7/2}$, see the fifth  column of \tref{tab-5d}. Among
these three contributions, the largest is from the $5d_{5/2}
- 6p_{3/2}$ transition.

Contributions to the tensor polarizabilities $\alpha_{2}(5d_{3/2})$ and
$\alpha_{2}(5d_{5/2})$  are given in columns three and six of \tref{tab-5d}.
The
$5d_{3/2} - 6p_{1/2}$ transition contributes 102\% of the
``Total'' shown on the last line of \tref{tab-5d} since two other
contributions ($5d_{3/2} - 6p_{3/2}$ and $5d_{3/2} -
5f_{5/2}$ partially cancel each other. A similar behavior is
found for $\alpha_{2}(5d_{5/2})$. The $5d_{5/2} - 6p_{3/2}$
transition contributes 87.6\% of the ``Total'' shown on the last
line of \tref{tab-5d}. An additional 10\% contribution is
from the $5d_{5/2} - 5f_{5/2}$ transition.
Contributions of transitions to highly exited states such as
$(9-26)p_{1/2}$, $(9-26)p_{3/2}$, $(10-26)f_{5/2}$ and
$(10-26)f_{7/2}$ are  small.

Contributions to the scalar polarizabilities of $6p$ levels
are shown in the second and fourth columns, respectively, of \tref{tab-6p}
and contributions to the tensor polarizability $\alpha_{2}(6p_{3/2})$
are given in column five of \tref{tab-6p}.

\section{Blackbody Radiatiion Shift}

Calculations of blackbody radiation shifts of clock frequencies in the monovalent ions Ca$^+$ and Sr$^+$  were
presented in
Refs.~\cite{mar-stark-07,mar-jpb-09,safr-ca-11,mar-bbr-12}.
The BBR shift for the $4s_{1/2}-3d_{5/2}$
transition in $^{43}$Ca$^+$ was calculated by
Arora {\it et al.\/} \cite{mar-stark-07} using the all-order SD method.
The SD method was also used by Jiang
{\it et al.\/} \cite{mar-jpb-09} to calculate the BBR shift of the $5s_{1/2}-4d_{5/2}$
clock transition in $^{88}$Sr$^+$.  A review of recent theoretical
calculations of BBR shifts in optical atomic clocks was presented by Safronova {\it et al.\/}
\cite{mar-bbr-12}.

The amplitude of the frequency-dependent electric field $E$
radiated by a black
body at temperature $T$ is given by the Planck radiation law
~\cite{mar-stark-07}. The frequency shift of an ionic state $v$ due to
such an electric field is related to the static scalar
polarizability $\alpha _{0}(v)$ of the state by
\begin{equation}
\Delta \nu =-\frac{1}{2}(831.9V/m)^{2}\left(
\frac{T(K)}{300}\right) ^{4}\alpha _{0}(v) \times (1+\eta )\label{eq-b1}
\end{equation}
where $\eta$ is a small ``dynamic'' correction
\cite{bbr-porsev,mar-bbr-sr}. We find that this dynamic
correction is negligible compared to the present 3\%
uncertainty of our calculated scalar polarizabilities.  The isotropic nature of the blackbody radiation field
leads to averaging out of the tensor
polarizability
effects. The  BBR shift of the clock transition frequency is the difference
between the BBR shifts of the states involved in the clock transition. The static BBR shift is
\begin{multline}\label{eq-b2}
\Delta (5d_{5/2}-6s_{1/2}) = \\
-\frac{1}{2}\left[\alpha _{0}(5d_{5/2}-6s_{1/2})\times
(831.9V/m)^{2}\left( \frac{T(K)}{300}\right) ^{4}\right]   .
\end{multline}
The evaluation of the BBR shift of the $5d_{5/2}-6s_{1/2}$ clock transition frequency in Lu$^{2+}$, therefore, involves accurate
calculations of static scalar polarizabilities of the $6s$
ground state and the $5d_{5/2}$ excited state. Here we use the
scalar polarizabilities
$\alpha_0(6s_{1/2})$ = 31.91$\pm$0.18 given in \tref{tab-6s} and
$\alpha_0(5d_{5/2})$ = 12.09$\pm$0.20 given in \tref{tab-5d}.
The resulting BBR shift of the clock transition
in Lu$^{2+}$ is
\[
\Delta _\text{BBR} (5d_{5/2}-6s_{1/2})\, =\, 0.1706 \pm 0.0023\ \text{Hz}.
\]
The $6s_{1/2}-5d_{5/2}$ transition frequency is $\nu_0$(Hz)=2.59$\times$10$^{14}$~Hz and
the ratio of the BBR shift to the transition frequency is 6.581$\times$10$^{-16}$ with an
uncertainty 0.089$\times$10$^{-16}$.

%31.91-12.09=19.82, SQRT[(0.18)**2 + (0.20)**]=0.269
%831.9**2=0.69206E+06
%0.69206E+06*2.48832E-08=0.017221
%0.5*(19.82(0.27))*0.017221=0.1706(0.0023)
%0.1706$\pm$0.0023/2.5925$\times$10$^{14}=6.5805\times$10$^{-16}
%0.0023=23E-4/2.5925$\times$10$^{14}=8.9E-18

In \tref{tab-bbr}, we present results for the dynamic
corrections to the BBR shift of the clock transition in
Lu$^{2+}$ at $T=300~K$. As mentioned earlier, the
final dynamic shift,
$\Delta \nu_{\rm BBR}^{\rm dyn}$($5d_{5/2}-6s_{1/2}$)= 0.000080~Hz,
is so small that it can be neglected compared to the uncertainty
induced by the polarizabilities ($\pm$0.0023~Hz).

\begin{table}
\caption{\label{tab-bbr} Dynamic correction to the BBR shift of
the $6s -5d$ clock transition in Lu$^{2+}$ at $T=300~K$ (in Hz). }
\begin{ruledtabular}
\begin{tabular}{lrcc}
 \multicolumn{1}{l}{}& \multicolumn{1}{c}{$\eta$} &
 \multicolumn{1}{c}{$\alpha_0(\omega=0)$} & \multicolumn{1}{c}{$\Delta \nu_{\rm BBR}^{\rm dyn}$}  \\  \hline     \\[-0.3pc]
 $6s_{1/2}-6p_{1/2}$    &0.000178  & &\\
 $6s_{1/2}-6p_{3/2}$    &0.000224   & & \\ [0.5pc]
      Total($6s_{1/2}$) &0.000402   & 31.91&  -0.000110\\ [0.5pc]

$5d_{5/2}-6p_{3/2}$     &0.000276 & & \\
$5d_{5/2}-5f_{5/2}$     &0.0000006 & & \\
$5d_{5/2}-5f_{7/2}$     &0.000012 & &  \\ [0.5pc]
Total($5d_{5/2}$)       &0.000288 & 12.09& -0.000030 \\
\hline
\multicolumn{3}{l}{Final\ $\Delta \nu_{\rm BBR}^{\rm dyn}$($5d_{5/2}-6s_{1/2}$)} &  0.000080\\
\end{tabular}
\end{ruledtabular}
\end{table}

%%%%%%%%%%%%%%%%%%%%%%%%%%%%%%%%%%%%%%%%%%%%%%%%%%%%%%%%%
\section{Hyperfine constants for $^{175}\text{Lu}^{2+}$}

Calculations of hyperfine constants follow the pattern described
earlier for calculations of transition matrix elements.
In \tref{tab-hyp}, we list hyperfine constants $A$ for
$^{175}$Lu$^{2+}$ and compare our values  with available
experimental measurements \cite{1971}.
In this table, we present the lowest-order $A^\text{(DF)}$ and
all-order $A^\text{(SD)}$ and $A^\text{(SDpT)}$
values for the $ns$, $np$ and $nd$ levels up to  $n$ = 8.
The nuclear spin and nuclear magnetic dipole moment of $^{175}$Lu$^{2+}$
used in these calculations are $I=7/2$ and $\mu= 2.2327~\mu_\text{N}$ \cite{NJS:05}.
Differences between $A^\text{(SD)}$ and
$A^\text{(SDpT)}$ are generally about 0.2\%, while the ratios $A^\text{(SD)}$
to $A^\text{(DF)}$  are  1.3-2.0 for some cases.
The largest difference between $A^\text{(SD)}$ and
$A^\text{(SDpT)}$ (about a factor of 2) occurs for the $5d_{5/2}$
level. For this state, even the signs of $A^\text{(DF)}$ and $A^\text{(SD)}$ are
different and the ratio of the magnitudes of $A^\text{(DF)}$ and $A^\text{(SD)}$ is
equal to 15. The difference between  $A^\text{(SD)}$ and
$A^\text{(MBPT)}$ for the $5d_{5/2}$ state is smaller
than between $A^\text{(SD)}$ and $A^\text{(SDpT)}$, emphasizing
the importance of higher-order correlation corrections for this transition.

We use the differences between
$A^\text{(SD)}$ and $A^\text{(SDpT)}$ to estimate the uncertainty
in our calculations.
Taking into account the
uncertainties in the experimental values~\cite{1971}, we conclude
that our results agree with the experimental values, which are given in the last
column of \tref{tab-hyp}. The largest  difference
between theory and experiment (9\%) is for the $6p_{1/2}$ level,
while the uncertainty given in Ref.~\cite{1971} for the $6p_{1/2}$ level is about 2\%.

Hyperfine constants $B$ (in MHz) for $^{175}$Lu$^{2+}$  are given in
  \tref{tab-hypb}.
The nuclear quadrupole moment of $^{175}$Lu is  $Q = 3.396$~b. \cite{NJS:16}.
Calculations of  $B^\text{(DF)}$,
$B^\text{(SD)}$,  $B^\text{(SDpT)}$ and $B^\text{(MBPT)}$  are listed in
  \tref{tab-hypb}.
The differences between $B^\text{(SD)}$ and $B^\text{(SDpT)}$ are in the range
0.1-0.5\% for all cases except the $5d$ levels where
$B^\text{(SD)}$ and $B^\text{(SDpT)}$ differ by about 2\%.
The ratios of the $B^\text{(SD)}$ and $B^\text{(SDpT)}$
are in the range 1.4 - 2.0  for all levels listed in
\tref{tab-hypb}, confirming the importance of the
correlation in the present calculations.
Values of $B^\text{(MBPT)}$ and $B^\text{(SDpT)}$ differ by about 2\%.

\begin{table}
\caption{\label{tab-hyp} Hyperfine constants $A$ (in MHz)  in
 $^{175}$Lu$^{2+}$ ($I$=7/2, $\mu$=2.2327~$\mu_N$   \cite{NJS:05}).
 Theoretical SD and SDpT results  are compared
with MBPT results and experimental data~\cite{1971}.}
\begin{ruledtabular}
\begin{tabular}{lrrrrr}
\multicolumn{1}{c}{Level} &
\multicolumn{1}{c}{$A^{(\rm DF)}$} &
\multicolumn{1}{c}{$A^{(\rm SD)}$} &
\multicolumn{1}{c}{$A^{(\rm SDpT)}$} &
\multicolumn{1}{c}{$A^{\rm (MBPT)}$}&
\multicolumn{1}{c}{$A^{(\rm expt.)}$} \\
\hline
   $6s_{1/2} $ &    10311.3 &   13158.4  &  13176.8& 13208.3& 13070(60)  \\
   $7s_{1/2} $ &     3466.9 &    4135.6  &   4140.2& 4137.3 & 4200(600)  \\
   $8s_{1/2} $ &     1610.3 &    1877.3  &   1879.3& 1876.2 & 1860(300)  \\
   $9s_{1/2} $ &      880.2 &    1012.4  &   1015.1& 1014.2 &            \\[0.4pc]

   $5d_{3/2} $ &      398.3 &     504.7  &    516.5& 543.3  &            \\
   $5d_{5/2} $ &      152.5 &     -10.2  &     -6.3& -11.3  &            \\
   $6d_{3/2} $ &       87.0 &     111.7  &    110.9& 114.8  &            \\
   $6d_{5/2} $ &       34.1 &      24.2  &     24.5&  25.0  &            \\
   $7d_{3/2} $ &       38.9 &      50.4  &     49.8&  50.5  &            \\
   $7d_{5/2} $ &       15.4 &      13.8  &     13.8&  13.6  &            \\[0.4pc]

   $6p_{1/2} $ &     2007.8 &    2778.3  &   2779.6&  2799.7  & 2520(60) \\
   $6p_{3/2} $ &      232.6 &     371.1  &    372.2&  360.1   & 390(90)  \\
   $7p_{1/2} $ &      786.7 &    1015.3  &   1014.2& 1027.3   &          \\
   $7p_{3/2} $ &       93.1 &     186.3  &    186.2&  137.9   &          \\
   $8p_{1/2} $ &      391.9 &     477.8  &    477.2&  500.0   &          \\
   $8p_{3/2} $ &       46.9 &     136.0  &    149.1&   68.6   &          \\
\end{tabular}
\end{ruledtabular}
\end{table}

 \begin{table}
\caption{\label{tab-hypb} Hyperfine constants $B$ (in MHz)  in
  $^{175}$Lu$^{2+}$.
The nuclear quadrupole moment of $^{175}$Lu$^{2+}$ is $Q = 3.396$~b \cite{NJS:16}.
Theoretical SD and SDpT results are compared with MBPT calculations.}
\begin{ruledtabular}
\begin{tabular}{lrrrr}
\multicolumn{1}{c}{Level} &
\multicolumn{1}{c}{$B^{(\rm DF)}$} &
\multicolumn{1}{c}{$B^{(\rm SD)}$} &
\multicolumn{1}{c}{$B^{(\rm SDpT)}$} &
\multicolumn{1}{c}{$B^{\rm (MBPT)}$}\\
\hline
   $5d_{3/2} $ &      1432.7&    1957.4&    2001.1&  2040.6\\
   $5d_{5/2} $ &      1692.4&    2456.4&    2509.8&  2499.5\\
   $6d_{3/2} $ &       312.9&     541.6&     540.2&   547.7\\
   $6d_{5/2} $ &       378.4&     696.4&     694.7&   700.2\\
   $7d_{3/2} $ &       140.1&     239.7&     238.1&   241.0\\
   $7d_{5/2} $ &       170.4&     308.3&     306.4&   308.4 \\[0.4pc]
   $6p_{3/2} $ &      2483.5&    3803.5&    3811.8&  3813.4\\
   $7p_{3/2} $ &       994.4&    1427.0&    1426.5&  1436.3\\
   $8p_{3/2} $ &       500.9&     645.7&     642.3&   689.2\\
\end{tabular}
\end{ruledtabular}
\end{table}

\section{Conclusion}
In summary, we have carried out a systematic high-precision
study of energies, transition matrix elements, polarizabilities, the blackbody shift and
hyperfine constants   for the $ns$,
$np$, $nd$ and $nf$ ($n \leq 9$), states of  Lu$^{2+}$ using a relativistic all-order
approach. Recommended values of the  atomic parameters are given along with
estimates of the corresponding uncertainties.
The theoretical energy values are in excellent
agreement with existing experimental data. Recommended values together with uncertainties are
provided for a large number of electric-dipole matrix elements.
Scalar and tensor
polarizabilities are evaluated for the ground state and low-lying
excited states in Lu$^{2+}$.  The  BBR shift of the $6s_{12}-5d_{5/2}$ clock transition frequency
in Lu$^{2+}$ ($\Delta _{\rm BBR} (6s_{1/2}-5d_{5/2})$) is found to be 0.1706$\pm$0.0023~Hz.
 $^{175}$Lu$^{2+}$.
Finally, the hyperfine constants $A$ and $B$ of $^{175}$Lu$^{2+}$ are determined for
low-lying levels up to $n$ = 9. This work provided recommended values of atomic properties of the ion Lu$^{2+}$,
critically evaluated for accuracy, in a systematic high-precision study.
These values may be useful for benchmark tests of theory, astrophysics applications
and for planning and analysis of various experiments that
depend on the level structure of the ion Lu$^{2+}$, including atomic clocks.

\acknowledgements
The authors owe a debt of gratitude to M. D. Barrett for helpful remarks on this manuscript. This work is partly supported by  NSF grant No. \ PHY-1520993.

%\bibliography{bbr-lu2}

\end{document}